\documentclass[conference]{IEEEtran}
\IEEEoverridecommandlockouts
\usepackage{amsmath, amssymb, cite, tikz, graphicx, xspace, mathrsfs, mathtools, psfrag, chemarrow, subfigure,color,soul,graphicx,mathtools,adjustbox}
\usepackage{kbordermatrix}
\usepackage{enumitem}
\usepackage{bbm}
\usepackage[utf8]{inputenc}
\usepackage{booktabs}
\usepackage{amscd}
\usepackage{amsmath}
\usepackage{amssymb}
\usepackage{amsthm}

\usepackage{epsfig}
\usepackage{verbatim}
\usepackage{graphicx}
\usepackage{amsthm}
\usepackage{color}
\usepackage{ multirow }
\usepackage[noend]{algpseudocode}
\usetikzlibrary{patterns}
\usetikzlibrary{angles} 
\usepackage{eucal}
\usepackage{algorithm}
\usepackage{tikz}
\usetikzlibrary{shapes, arrows, decorations.markings, arrows.meta}
\usetikzlibrary{patterns} 
\usetikzlibrary{arrows,automata,positioning,chains,calc}
\usetikzlibrary{quotes,angles}

\makeatletter
\def\BState{\State\hskip-\ALG@thistlm}
\makeatother
\newtheorem{theorem}{Theorem}

\newtheorem{proposition}{Proposition}
\newtheorem{definition}{Definition}
\newcommand{\D}{\Delta}
\newcommand{\at}{\alpha_{t}}
\newcommand{\Dm}{\Delta_{\text{max}}}
\newcommand{\Em}{E_{\text{max}}}
\newcommand{\ps}{p_{\text{s}}}

\newcommand{\Bb}{\overline{\beta}}
\DeclareMathOperator*{\argmin}{arg\,min}

\begin{document}
\setlength{\abovecaptionskip}{-3pt}
\setlength{\belowcaptionskip}{1pt}
\setlength{\floatsep}{1ex}
\setlength{\textfloatsep}{1ex}

\title{Optimizing Version Innovation Age for Monitoring Markovian Source in Energy-Harvesting Systems}
\author{
	Mehrdad Salimnejad, 
 Anthony Ephremides,
		Marios Kountouris,  
  and Nikolaos Pappas\thanks{\scriptsize M. Salimnejad and N. Pappas are with the Department of Computer and Information Science Linköping University, Sweden, email: \{\texttt{mehrdad.salimnejad, nikolaos.pappas\}@liu.se}. A. Ephremides is with the Electrical and Computer Engineering, University of Maryland, College Park, MD, USA, email: \texttt{etony@umd.edu}. M. Kountouris is with the Department of Computer Science and Artificial Intelligence, Andalusian Research Institute in Data Science and Computational Intelligence (DaSCI), University of Granada, Spain, email: \texttt{mariosk@ugr.es}.}}
\maketitle
\begin{abstract}
We study the real-time remote tracking of a two-state Markov process by an energy harvesting source. The source decides whether to transmit over an unreliable channel based on the state. We formulate this scenario as a Markov decision process (MDP) to determine the optimal transmission policy that minimizes the average Version Innovation Age (VIA) as a performance metric. We demonstrate that the optimal transmission policy is threshold-based, determined by the battery level, source state, and VIA value. We numerically verify the analytical structure of the optimal policy and compare the performance of our proposed policy against two baseline policies across various system parameters, establishing the superior performance of our approach.
\end{abstract}
\vspace{-0.15cm}
\section{Introduction}
\vspace{-0.1cm}
\par Timely and efficient information exchange has emerged as an important area of research in communication systems designed for time-critical applications \cite{abd2019role, shreedhar2019age}. In such systems, sensors continuously monitor physical processes and transmit status updates via a communication network to a receiver, enabling further processing and decision-making. The reliability and accuracy required for effective decision-making in these time-sensitive and data-intensive systems significantly depend on \emph{the freshness of the information}. Nevertheless, resource limitations, such as limited bandwidth, energy restrictions, unreliable or intermittent channels, and other factors, pose significant challenges. Therefore, designing optimal strategies for data generation, transmission, and processing in resource-constrained networks is of cardinal importance. These stringent requirements have spurred the development of \emph{goal-oriented semantics-empowered communications} \cite{kountouris2021semantics,popovski2020semantic,popovski2022perspective}. This emerging paradigm highlights the importance of information utility and introduces innovative strategies for timely data generation, transmission, and utilization to effectively achieve specific goals within status update systems. A set of semantics-aware metrics that quantify the timeliness or freshness and the significance of information has been proposed \cite{kaul2012real,maatouk2020age,pappas2021goal,MSalimnejadTCOM2024,yates2021age}. A new semantics-aware metric, termed \emph{Version Innovation Age (VIA)}, was recently introduced in \cite{salimnejad2024age}. This metric measures the number of outdated information versions at the receiver compared to those at the source when the source is in a \emph{specific state}. VIA differs from the Version Age of Information (VAoI) \cite{yates2021age}, which focuses on the lag in version updates between the receiver and the source. Unlike VAoI, VIA focuses on the importance of updates based on the source's state rather than their frequency. A key challenge in this field is optimizing semantics-aware metrics while accounting for resource constraints in status update systems. In practice, these systems are powered by batteries with limited lifespans and rely on energy harvesting (EH) technologies. Therefore, optimizing the management of stored energy to ensure efficient sampling and transmission of status updates is essential. Several studies have investigated the optimization of semantics-aware metrics in the context of energy-constrained networks \cite{Yates2015,wu2017optimal,arafa2019age,abd2019online, Stamatakis2019,hatami2022demand,DelfaniWiopt2023}. The work in \cite{Yates2015} studied a scenario where a source sends status updates to a service facility powered by an unpredictable EH system and showed that for sources with large batteries, the optimal policy to minimize the average Age of Information (AoI) is to delay sending updates until previous ones have been fully processed. In \cite{wu2017optimal,arafa2019age,abd2019online}, the authors investigated real-time remote sensing under energy constraints and proposed optimal online update policies tailored to battery-size variations. In \cite{Stamatakis2019}, the authors derived optimal transmission policies for an EH system monitoring a stochastic process. The problem was modeled as a Markov decision process (MDP) with a cost function incorporating linear and nonlinear penalties, utilizing two AoI variables. The study in \cite{hatami2022demand} aimed to minimize on-demand AoI in EH multi-sensor networks using an MDP model, proposing an optimal iterative algorithm and a simpler sub-optimal one for large-scale sensor networks. \cite{DelfaniWiopt2023} optimized the average VAoI in a real-time monitoring system with an EH sensor, using an MDP framework to determine the best update policy from either stored or fresh transmissions.
\par This paper investigates a time-slotted communication system consisting of a source, an EH sensor, and a receiver. In each time slot, the EH sensor observes status updates generated by a two-state Markov source and decides whether to transmit these updates as packets over an unreliable channel. We use the VIA as the performance metric and aim to design a transmission policy that minimizes the average VIA while accounting for the energy budget. To this end, we leverage an MDP framework to derive the optimal transmission policy. Furthermore, we examine the structural properties of the optimal solution and assess the impact of various system parameters using analytical and numerical methods.
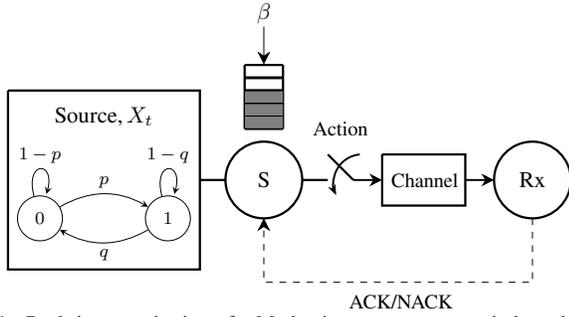
\begin{figure}[!t]
	\centering
 \resizebox{0.42\textwidth}{!}{
\begin{tikzpicture}[start chain=going left,>=stealth,node distance=2cm,on grid,auto]
	\footnotesize
	\node[state, on chain]               at (0,-2)  (2) {$1$};
	\node[state, on chain]              at  (0,-2)   (1) {$0$};
	\path[->]
	(1)   edge[loop above] node  {$1-p$}   (1)
	(1) edge  [bend left=30] node {$p$} (2)
	
	(2) edge  [bend left=30] node {$q$} (1)
	(2) edge  [loop above] node {$1-q$} (2)
	;
	\draw [line width=0.35mm](-2.5,0)--(0.5,0)--(0.5,-2.8)--(-2.5,-2.8)--(-2.5,0);
	\node(a) at (-1,-0.4)  {\normalsize{{Source}},\! $X_{t}$};
	\draw [line width=0.35mm](1.5,-1.4) circle (0.6cm);
	\node(b) at (1.5,-1.4)  {\normalsize	\text{S}};
	\draw [line width=0.35mm](2.1,-1.4) -- (2.5,-1.4);
	\draw [line width=0.35mm](0.5,-1.4) -- (0.9,-1.4);
	\draw [line width=0.35mm](2.5,-1.)to(2.9,-1.4);
	\draw[line width=0.3mm,-{Stealth[length=1.7mm]}] (3,-1.) arc (95:205:4.5mm) ;
	\draw [line width=0.3mm,-{Stealth[length=2mm, width=2mm]}](2.9,-1.4)--(3.35,-1.4);
	\node(a1) at (2.7,-0.6)  {\small	\text{Action}};
	\node(b1) at (4.02,-1.4)  {\small	\text{Channel}};
	\draw [line width=0.35mm](3.35,-1.)--(4.65,-1.)--(4.65,-1.8)--(3.35,-1.8)--(3.35,-1);
	\draw [line width=0.3mm,-{Stealth[length=2mm, width=2mm]}](4.65,-1.4) -- (5.1,-1.4);
	\node(b2) at (3.7,-3.3)  {\small	\text{ACK/NACK }};
	\draw [line width=0.35mm](5.7,-1.4) circle (0.6cm);
	\node(c) at (5.7,-1.4)  {\normalsize	\text{Rx}};
	\draw [-{Stealth[length=2mm, width=2mm]}](1.5,1) -- (1.5,0.41);
	\node(c1) at (1.5,1.2)  {\normalsize	$\beta$};
	\draw [line width=0.35mm](1.2,0.4) -- (1.2,-0.6);
	\draw [line width=0.35mm](1.2,0.4) -- (1.8,0.4);
	\draw [line width=0.35mm](1.8,0.4) -- (1.8,-0.6);
	\draw [line width=0.35mm](1.8,-0.6) -- (1.2,-0.6);
	\draw [line width=0.45mm](1.2,0.2) -- (1.8,0.2);
	\draw [line width=0.45mm](1.2,0) -- (1.8,0);
	\draw [line width=0.45mm](1.2,-0.2) -- (1.8,-0.2);
	\draw [line width=0.45mm](1.2,-0.4) -- (1.8,-0.4);
\filldraw [line width=0.1mm,gray](1.21,-0.02) -- (1.79,-0.02)--(1.79,-0.18)--(1.21,-0.18);
\filldraw [line width=0.1mm,gray](1.21,-0.22) -- (1.79,-0.22)--(1.79,-0.38)--(1.21,-0.38);
\filldraw [line width=0.1mm,gray](1.21,-0.42) -- (1.79,-0.42)--(1.79,-0.58)--(1.21,-0.58);
	\draw [-{Stealth[length=2mm, width=2mm]},dashed](5.7,-2) -- (5.7,-3)--(1.5,-3)--(1.5,-2);
\end{tikzpicture}}
	\vspace{-0.1cm}
	\caption{Real-time monitoring of a Markovian source over a wireless channel.}
	\label{system_model_fig}
\end{figure}  
\vspace{-0.2cm}
\section{System Model}
\vspace{-0.1cm}
\par We consider a time-slotted communication system where an EH sensor monitors an information source and transmits status updates to a receiver responsible for performing an action, as shown in Fig. \ref{system_model_fig}. The information source at time slot $t$, denoted as $X_{t}$, is represented as a two-state discrete-time Markov chain (DTMC) $\{X_{t}, t \in \mathbb{N}\}$. Therein, the state transition probability $\mathrm{Pr}\big[X_{t+1}=j \big|X_{t}=i\big]$ represents the probability of transitioning from state $i$ to $j$ and can be defined as $\mathrm{Pr}\big[X_{t+1}=j\big|X_{t}=i\big] = \mathbbm{1} (i=0,j=0)(1-p)+\mathbbm{1}(i=0,j=1)p+\mathbbm{1}(i=1,j=0)q+\mathbbm{1}(i=1,j=1)(1-q)$, where $\mathbbm {1}(\cdot)$ is the indicator function. The sensor has a finite-capacity buffer that can store a maximum of $E_{\text{max}}$ energy units. At each time slot, the energy arrival process, denoted by $b_{t} \in \{0, 1\}$, is modeled as a Bernoulli process with an average probability $0 < \beta < 1$. This means that $\mathrm{Pr}[b_{t} = 1] = \beta$ and $\mathrm{Pr}[b_{t} = 0] = \Bb=1 - \beta$. We define $\alpha_{t} \in \{0,1\}$ as the decision of the sensor node at time slot $t$ to either transmit the status updates $(\alpha_{t} = 1)$ or remain idle $(\alpha_{t} = 0)$. We assume that each transmission action consumes one unit of energy from the battery, and transmission does not occur if the battery is empty. Therefore, the battery level at time slot $t$, denoted by $e_{t} \in \{0,1,\cdots,\Em\}$, evolves as $e_{t+1}=\min\{e_{t}+b_{t}-\at,\Em\}$.
It is assumed that each transmission occurs over a wireless channel, and the channel state $h_{t}$ equals $1$ if the information is transmitted and successfully decoded by the receiver and $0$ otherwise. We define the success and failure probabilities as $\ps = \mathrm{Pr}[h_{t}=1]$ and $p_{f} = \mathrm{Pr}[h_{t}=0] = 1 - \ps$, respectively. Acknowledgment (ACK) and negative acknowledgment (NACK) packets are employed to inform the transmitter about the success or failure of transmissions, with the assumption that these packets are delivered immediately and error-free. This paper considers the VIA as the performance metric, which measures the number of outdated versions at the receiver compared to the source when the source is in a specific state \cite{salimnejad2024age}. Let $\Delta_{t}$ represent the VIA at time slot $t$. To enable real-time monitoring of the source, at the beginning of each time slot, the sensor measures the battery level and, based on the state of the source and $\Delta_{t}$, decides whether to transmit the status updates to the receiver. We define the evolution of $\Delta_{t}$ as follows:
\vspace{-0.2cm}
\begin{align}
	\label{VIA}
	\!\D_{t+1}\!\!=\!\!
	\begin{cases}
		\D_{t}, & \!\!\!\!\parbox[t]{9cm}{{$X_{t+1}= X_{t}$ \!\text{and} $\{\at\!=\!0,\\ \text{or} \hspace{0.1cm} (\at\!=\!1, h_{t}\!=\!0)\}$,}}\\
		\!\!\min\{\D_{t}\!+\!1,\Dm\}, & \!\!\!\!\parbox[t]{9cm}{{$X_{t+1}\neq X_{t}$ \!\text{and} $\{\at\!=\!0,\\ \text{or} \hspace{0.1cm} (\at\!=\!1, h_{t}\!=\!0)\}$,}}\\
		1, &\!\!\! \parbox[t]{9cm}{$X_{t+1}\!\neq\! X_{t}$, $\at \!=\!1, h_{t}\!=\!1$}\\
		0,&\!\!\! \parbox[t]{9cm}{$X_{t+1}\!=\! X_{t}$, $\at \!=\!1, h_{t}\!=\!1$}
	\end{cases}
\end{align}
where $\Delta_{\text{max}}$ is the maximum value of the VIA.
\vspace{-0.2cm}
\section{Problem Formulation}
\label{MDPProblemSection1}
\vspace{-0.12cm}
\par In this section, we obtain the optimal transmission policy that minimizes the average VIA. The transmission policy, denoted by $\pi$, is defined as a sequence of actions $\pi = \big(\alpha^{\pi}_{1}, \alpha^{\pi}_{2}, \cdots\big)$ where $\alpha^{\pi}_{t}=1$ indicates the sensor transmits the status updates at time slot $t$; otherwise, $\alpha^{\pi}_{t}=0$. By defining $\Pi$ as the set of all possible causal policies, our problem can be defined as an infinite horizon average cost MDP as follows
\begin{itemize}
    \item \textbf{States}: The state of the system at time slot $t$ is defined by $S_{t} = \big(e_{t}, X_{t}, \Delta_{t}\big)$, where $e_{t} \in \{0, 1, \cdots, \Em\}$ represents the battery level, $X_{t}$ denotes the state of the information source, and $\Delta_{t}$ is the VIA.
    \item \textbf{Actions}: The action at time slot $t$, denoted by $\at$, indicates whether the sensor S decides to transmit $(\at=1)$ or remain idle $(\at=0)$.
     \item \textbf{Transition Probabilities}: The transition probabilities from $S_{t}$ to $S_{t+1}$ under the action $\at$ are defined in Section \ref{TransitionProb}.
     \item \textbf{Cost}: The cost of the MDP problem at time slot $t$, denoted by $C\big(S_{t}, \at\big)$, is the VIA $\Delta_{t}$.
\end{itemize}
Now, we formulate our MDP problem as follows
\vspace{-0.1cm}
\begin{align}
    \label{MDPProblem}
    \underset{\pi\in \Pi}{\text{min}} \limsup_{T\to \infty} \frac{1}{T}E\left[\sum_{t=1}^{T} \Delta^{\pi}_{t}\Big| S_{1}=s\right].
\end{align}

\subsection{Transition Probabilities}
\label{TransitionProb}
\par The transition probabilities from state $S_{t} = \big(e_{t}, X_{t}, \Delta_{t}\big)$ to state $S_{t+1} = \big(e_{t+1}, X_{t+1}, \Delta_{t+1}\big)$ under action $\at$ are defined as $\mathrm{Pr}\big[S_{t+1}\big|S_{t}, \at \big]$. To derive these transition probabilities, based on the battery level at time slot $t$, we consider the following two different cases:
\begin{enumerate}
    \item $e_{t}\!=\!0$: when the battery at time slot $t$ is empty, sensor S does not transmit the status updates, therefore $\mathrm{Pr}\big[S_{t+1}\big|S_{t},\at\!=\!0\big]= \mathrm{Pr}\big[S_{t+1}\big|S_{t},\at\!=\!1\big]$. In this case,  $\Delta_{t+1}$ increases by one if the source's state changes; otherwise, $\Delta_{t+1}$ remains in its previous state, i.e., $\Delta_{t+1} = \Delta_{t}$. Now, depending on the energy arrival process at time slot $t+1$, we can write the transition probabilities as follows
    \vspace{-0.5cm}
    \begin{align}
    \label{TransProb_001i}
&\mathrm{Pr}\big[S_{t+1}\big|S_{t},\at=0\big]= \mathrm{Pr}\big[S_{t+1}\big|S_{t},\at=1\big]=\notag\\
	&\begin{cases}
		\Bb(1-p), &\!\!\!\!\!\parbox[t]{9cm}{{$X_{t} \!=\! 0$, $\D_{t} \!=\! i$, $e_{t+1}\!=\!e_{t}$,$X_{t+1}\!=\!0$,$\D_{t+1}\!=\!i$,}}\\
		\beta(1-p),&\!\!\!\!\!\parbox[t]{9cm}{{$X_{t} \!=\! 0$, $\D_{t} \!=\! i$, $e_{t+1}\!=\!\min\{e_{t}+1,\Em\}$,\\
		$X_{t+1}\!=\!0$, $\D_{t+1}\!=\!i$,}}\\
		\Bb p,&\!\!\!\!\!\parbox[t]{9cm}{{$X_{t} = 0$, $\D_{t} = i$, $e_{t+1}=e_{t}$, $X_{t+1}=1$,\\$\D_{t+1}=\min\{i+1,\Dm\}$,}}\\
		\beta p,&\!\!\!\!\!\parbox[t]{9cm}{{$X_{t} = 0$, $\D_{t} = i$, $e_{t+1}\!=\!\min\{e_{t}\!+\!1,\Em\}$,\\$X_{t+1}=1$, $\D_{t+1}=\min\{i+1,\Dm\}$,}}\\
  \Bb (1-q), &\!\!\!\!\!\parbox[t]{9cm}{{$X_{t} \!\!=\!\! 1$, $\D_{t} \!\!=\!\! i$, $e_{t+1}\!\!=\!\!e_{t}$, $X_{t+1}\!\!=\!\!1$, $\D_{t+1}\!=\!i$,}}\\
		\beta(1-q),&\!\!\!\!\!\parbox[t]{9cm}{{$X_{t} \!=\! 1$, $\D_{t} \!=\! i$, $e_{t+1}\!=\!\min\{e_{t}+1,\Em\}$,\\
		$X_{t+1}=1$, $\D_{t+1}=i$,}}\\
		\Bb q,&\!\!\!\!\!\parbox[t]{9cm}{{$X_{t} = 1$, $\D_{t} = i$, $e_{t+1}=e_{t}$, $X_{t+1}=0$,\\
				$\D_{t+1}=\min\{i+1,\Dm\}$,}}\\
		\beta q,&\!\!\!\!\!\parbox[t]{9cm}{{$X_{t} \!=\! 1$, $\D_{t} \!=\! i$, $e_{t+1}\!=\!\min\{e_{t}+1,\Em\}$,\\
		$X_{t+1}=0$, $\D_{t+1}=\min\{i+1,\Dm\}$.}}.
	\end{cases}
\end{align}
\vspace{-0.45cm}
\item $e_{t}\neq 0$: when the battery level at time slot $t$ is non-empty, the transition probabilities for $\at=0$ are given by \eqref{TransProb_001i}. Furthermore, for $\at=1$, two cases are considered based on the state of the VIA $\Delta_{t}$. First, if $\Delta_{t} = 0$, then $\Delta_{t+1} = 0$ if the source state remains unchanged; otherwise, $\Delta_{t+1} = 1$. In this scenario, the success or failure of the transmission at time slot $t$ does not impact $\Delta_{t+1}$. The transition probabilities can be derived in a manner similar to \eqref{TransProb_001i}, but with the key difference that, depending on the energy arrival process at time slot $t+1$, the energy level $e_{t+1}$ can be either $e_{t}$ or $e_{t} - 1$. Now, we assume $\Delta_{t} = i>0$. In this scenario, we consider two cases based on whether the transmission succeeds or fails.
First, the transmission succeeds at time slot $t$. In this case, when the source state remains unchanged, $\Delta_{t+1} = 0$, otherwise, $\Delta_{t+1} = 1$. Furthermore,  there is a failure the transmission at time slot $t$.  In this case, when the source state remains unchanged, $\Delta_{t+1} = i$, otherwise, $\Delta_{t+1} = i+1$. Now, the transition probabilities when $X_{t} = 0$ are obtained as follows. Similarly, the transition probabilities for $X_{t} = 1$ can be derived.
\end{enumerate}
\vspace{-0.1cm}
\begin{align}
\label{TranProb_ED}
&P\big[S_{t+1}\big|S_{t},\at=1\big]=\notag\\
	&\!\!\begin{cases}
		\beta(1-p)\ps, &\!\!\!\!\!X_{t} \!=\! 0, \Delta_{t} \!=\! i, e_{t+1}\!=\!e_{t},X_{t+1}\!=\!0,\D_{t+1}\!=\!0,\\
		\Bb (1-p)\ps,&\!\!\!\!\!X_{t} \!=\! 0, \Delta_{t} \!=\! i, e_{t+1}\!=\!e_{t}\!-\!1,X_{t+1}\!=\!0,\D_{t+1}\!=\!0,\\
  \beta(1-p)p_{f}, &\!\!\!\!\!X_{t} \!=\! 0, \Delta_{t} \!=\! i, e_{t+1}\!=\!e_{t},X_{t+1}\!=\!0,\D_{t+1}\!=\!i,\\
  \Bb (1-p)p_{f},&\!\!\!\!\!X_{t} \!=\! 0, \Delta_{t} \!=\! i, e_{t+1}\!=\!e_{t}\!-\!1,X_{t+1}\!=\!0,\D_{t+1}\!=\!i,\\
		\beta p\ps,&\!\!\!\!\!X_{t} \!=\! 0, \Delta_{t} \!=\! i, e_{t+1}\!=\!e_{t},X_{t+1}\!=\!1,\D_{t+1}\!=\!1,\\
		\Bb p\ps,&\!\!\!\!\!X_{t} \!=\! 0, \Delta_{t} \!=\! i, e_{t+1}\!=\!e_{t}\!-\!1,X_{t+1}\!=\!1,\D_{t+1}\!=\!1,\\
		\beta p p_{f},&\!\!\!\!\!\parbox[t]{9cm}{{$X_{t} = 0$, $\Delta_{t} = i$, $e_{t+1}=e_{t}$, $X_{t+1}=1$,\\$\D_{t+1}=\min\{i+1,\Dm\}$,}}\\
		\Bb pp_{f},&\!\!\!\!\!\parbox[t]{9cm}{{$X_{t} = 0$, $\Delta_{t} = i$, $e_{t+1}=e_{t}-1$, $X_{t+1}=1$,\\$\D_{t+1}=\min\{i+1,\Dm\}$.}}
	\end{cases}
\end{align}
\subsection{Analytical Results}
\vspace{-0.1cm}
\par We here present the analytical results regarding the optimal transmission policy for the MDP problem in (\ref{MDPProblem}).
\vspace{-0.1cm}
\begin{definition}
		An MDP is considered weakly accessible if its states can be divided into two subsets, $\mathfrak{S}_{a}$ and $\mathfrak{S}_{b}$. All states in $\mathfrak{S}_{a}$ are transient under any stationary policy, and for every state $S$ and $S'$ in $\mathfrak{S}_{b}$, state $S'$ can be reached from state $S$.
	\end{definition}
	\vspace{-0.3cm}
	\begin{proposition}
 \label{prop1}
		The MDP problem in \eqref{MDPProblem} is weakly accessible.
	\end{proposition}
 \vspace{-0.15cm}
 \begin{IEEEproof}
    We demonstrate that, under a stationary stochastic policy $\pi$, any state $S'\!=\! (e',X',\Delta')$ can be reached from any other state $S(e,X,\Delta)$, where the action $a \!\in\! \{0, 1\}$ at each state is selected randomly with a positive probability. In particular, the state $e' \!<\! e$ is reachable from $e$ with positive probability by performing action $a\! =\! 1$ for $(e \!-\! e')$ time slots. Similarly, the state $e'\! \geqslant \!e$ can be reached from $e$ with positive probability by taking action $a = 0$ for $(e' - e)$ time slots. Once the system reaches the battery state $e'$, the battery state remains the same with a positive probability, regardless of future actions. Consequently, for the remainder of the proof, we will consider the battery state to be $e'$. Furthermore, the state $\Delta' \!<\! \Delta$ can be achieved from $\Delta$ with a positive probability by taking action $a \!=\! 1$ for one time slot (with no change in the source state), followed by $\Delta'$ time slots of action $a\!=\!0$ and the source state changing $\Delta'$ times. On the other hand, the state $\Delta' \!\geqslant\! \Delta$ can be reached from $\Delta$ by executing action $a\!=\!0$ for $\Delta' \!-\! \Delta$ slots and having the source state change $\Delta' \!-\! \Delta$ times.
 \end{IEEEproof}
 \begin{proposition}
 \label{prop2}
      In the MDP problem described in \eqref{MDPProblem}, the optimal policy $\pi^{*}$ results in the same optimal average cost $\theta^{*}$ for all initial states, and it satisfies the Bellman's equation at the state $S = (e, X, \Delta)$ as follows
       \vspace{-0.2cm}
      \begin{align}
      \label{BellmanEquation}
         \theta^{*}+ V(S) &= \underset{a\in \{0,1\}}{\min} \Biggl\{\Delta+\sum_{S^{\prime}\in \mathfrak{S}}\mathrm{Pr}\big[S^{\prime}|S,a\big]V(S^{\prime})\Biggr\},
      \end{align}
       \vspace{-0.3cm}
      \begin{align}
      \label{OptimalPolicy}
          \pi^{*}(S) \in  \argmin_{a\in\left\{0,1\right\}}\Biggl\{\Delta+\sum_{S^{\prime}\in \mathfrak{S}}\mathrm{Pr}\big[S^{\prime}|S,a\big]V(S^{\prime})\Biggr\},
      \end{align}
      where $V(S)$ is the value function of the MDP problem.
 \end{proposition}
 \begin{IEEEproof}
According to Proposition 1, problem \eqref{MDPProblem} is weakly accessible. Consequently, Proposition 4.2.3 in \cite{bertsekas2011dynamic} ensures that the optimal average cost is the same for all initial states. Furthermore, Proposition 4.2.6 in \cite{bertsekas2011dynamic} confirms the existence of an optimal policy, and Proposition 4.2.1 in \cite{bertsekas2011dynamic} states that if we can find $\theta^{*}$ and $V(S)$ that satisfy \eqref{BellmanEquation}, the optimal policy is given by \eqref{OptimalPolicy}. According to \eqref{OptimalPolicy}, the optimal policy $\pi^{*}(S)$ depends on $V(\cdot)$, which generally cannot be solved in closed form \cite{bertsekas2011dynamic}. Various numerical algorithms, such as value iteration and policy iteration algorithms, can be used to solve \eqref{OptimalPolicy}.
 \end{IEEEproof}
 \begin{theorem}
 \label{theorem_optimalpolicy}
	The optimal policy of the MDP problem \eqref{MDPProblem} is a threshold policy.
\end{theorem}
\begin{IEEEproof}
See Appendix \ref{Appendix1: theorem1}.
\end{IEEEproof}
\section{Numerical Results}
\par In this section, we numerically study the structural properties of the optimal transmission policy derived from the value iteration algorithm. We evaluate the performance of the average VIA under varying system parameters and confirm that the optimal transmission policy exhibits a threshold-based structure. For comparison purposes, we adopt two additional baseline policies, namely the \emph{Randomized Stationary (RS)} and the \emph{Greedy policies}. In the RS policy, the transmitter sends status updates with a probability of $p_{\alpha}$ when the battery is not empty. Here we assume that $p_{\alpha} \!=\! 0.5$. In the greedy policy, the transmitter sends status updates whenever the battery is not empty. In both policies, transmissions occur regardless of the state of the source and VIA. Furthermore, the simulation results are obtained by averaging over $10^7$ time slots.
\par
\begin{figure}[!t]
		\centering
		\subfigure[$X = 0, \beta = 0.2$ ]{\includegraphics[trim=0.5cm 0.05cm 1.1cm 0.6cm,width=0.48\linewidth, clip]{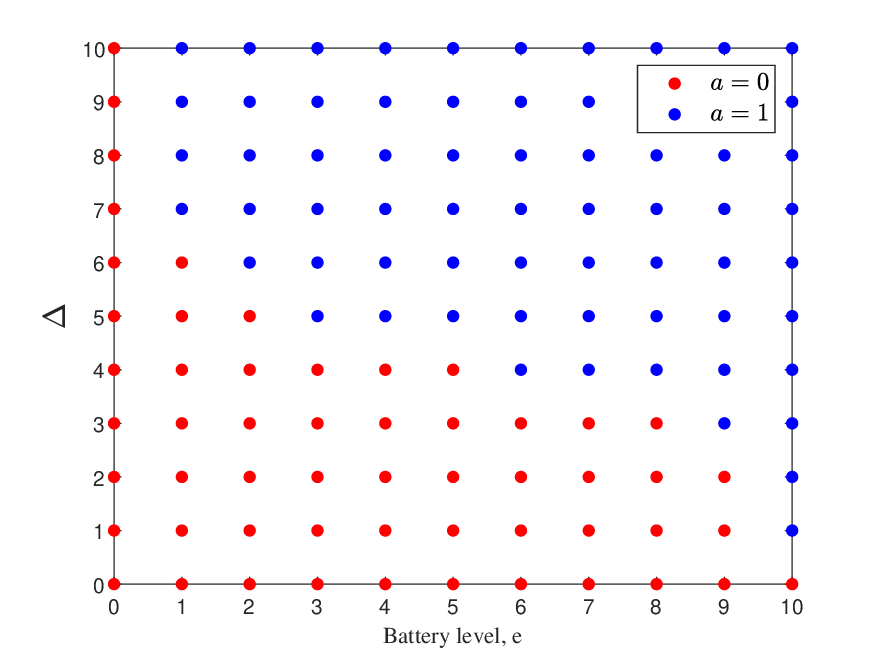}
		}
		\subfigure[$X =1, \beta = 0.2$]{\centering
			\includegraphics[trim=0.5cm 0.05cm 1.1cm 0.6cm,width=0.48\linewidth, clip]{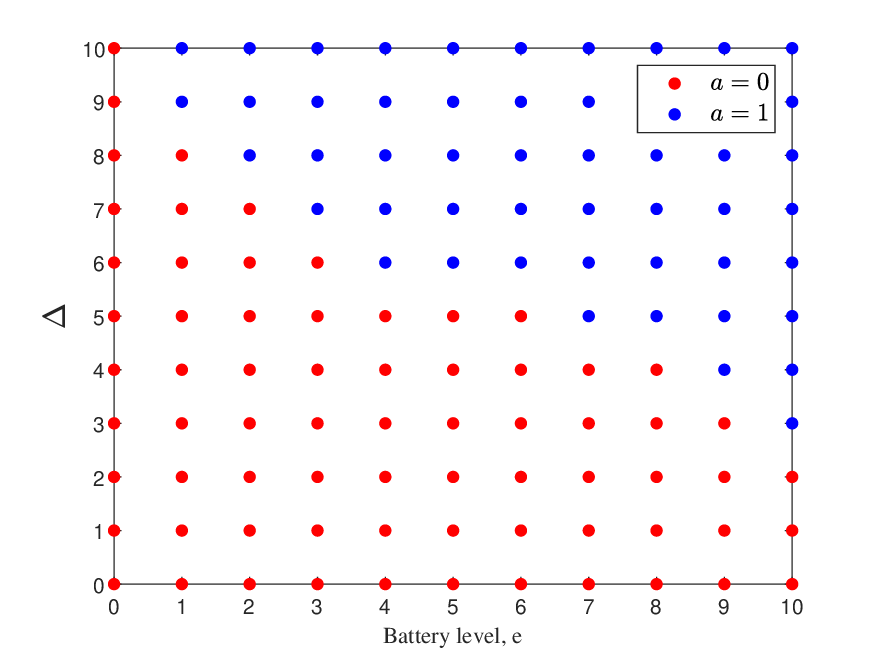}
		}
  \subfigure[$X =0, \beta = 0.4$]{\centering
			\includegraphics[trim=0.5cm 0.05cm 1.1cm 0.6cm,width=0.48\linewidth, clip]{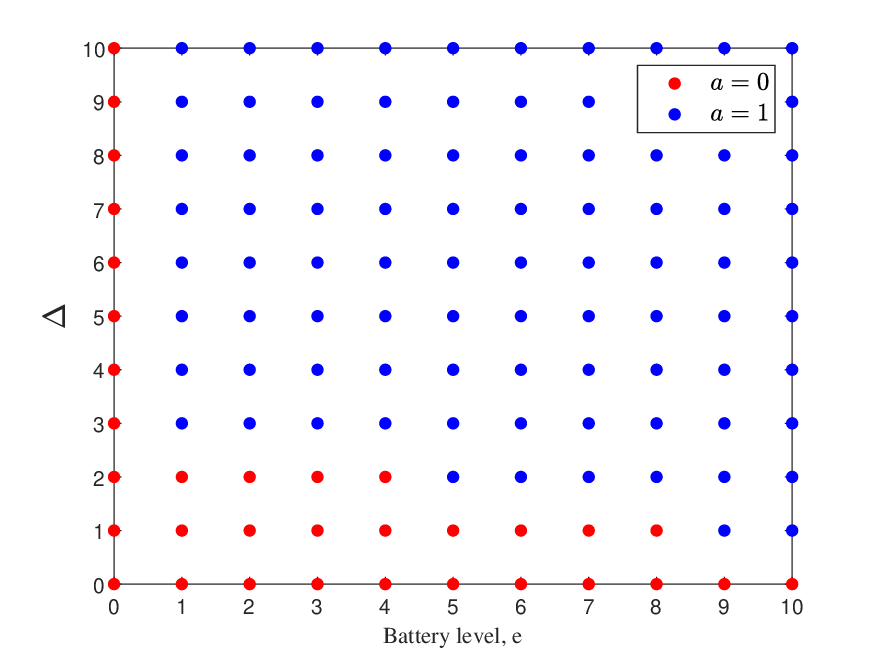}
		}
  \subfigure[$X =1, \beta = 0.4$]{\centering
			\includegraphics[trim=0.5cm 0.05cm 1.1cm 0.6cm,width=0.48\linewidth, clip]{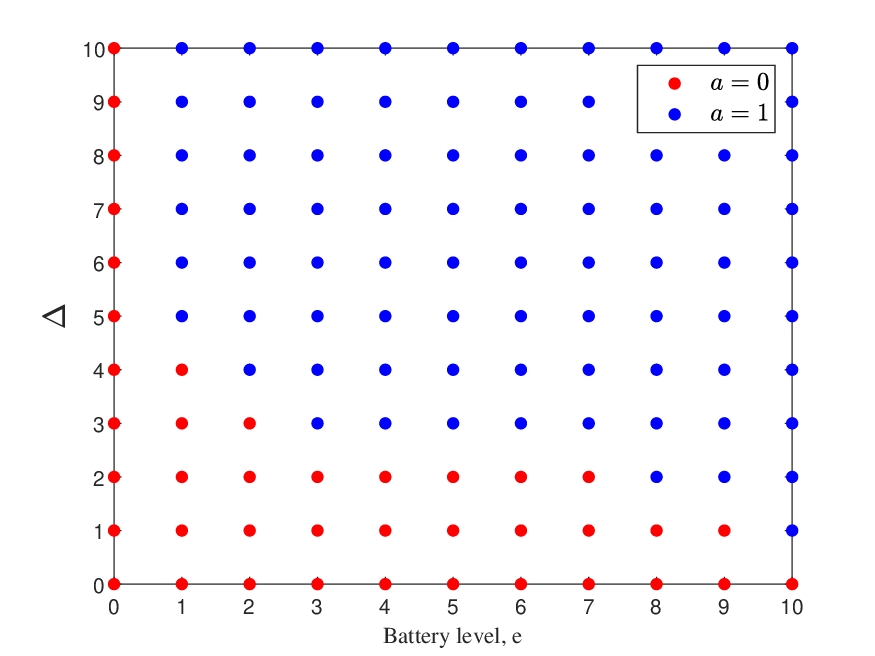}
		}
\caption{The structure of the optimal transmission policy for $p = 0.4$, $q = 0.7$, and $p_{s} = 0.5$.}
\label{VIA_Battery_p0.4q0.7}
\end{figure}

\begin{figure}[!t]
		\centering 
		\subfigure[$X = 0, \beta = 0.2$ ]{\includegraphics[trim=0.5cm 0.05cm 1.1cm 0.6cm,width=0.48\linewidth, clip]{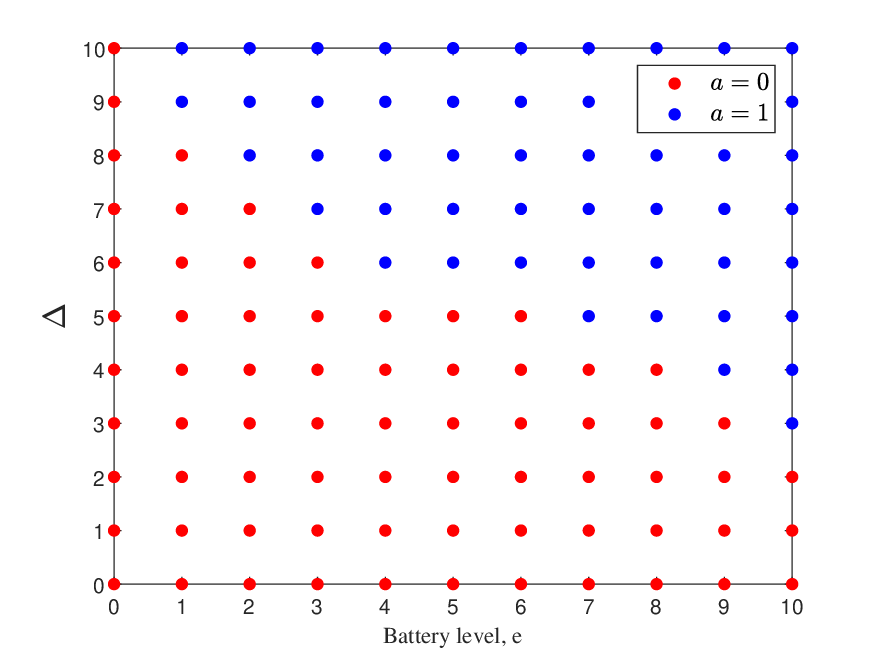}
		}
		\subfigure[$X =1, \beta = 0.2$]{\centering
			\includegraphics[trim=0.5cm 0.05cm 1.1cm 0.6cm,width=0.48\linewidth, clip]{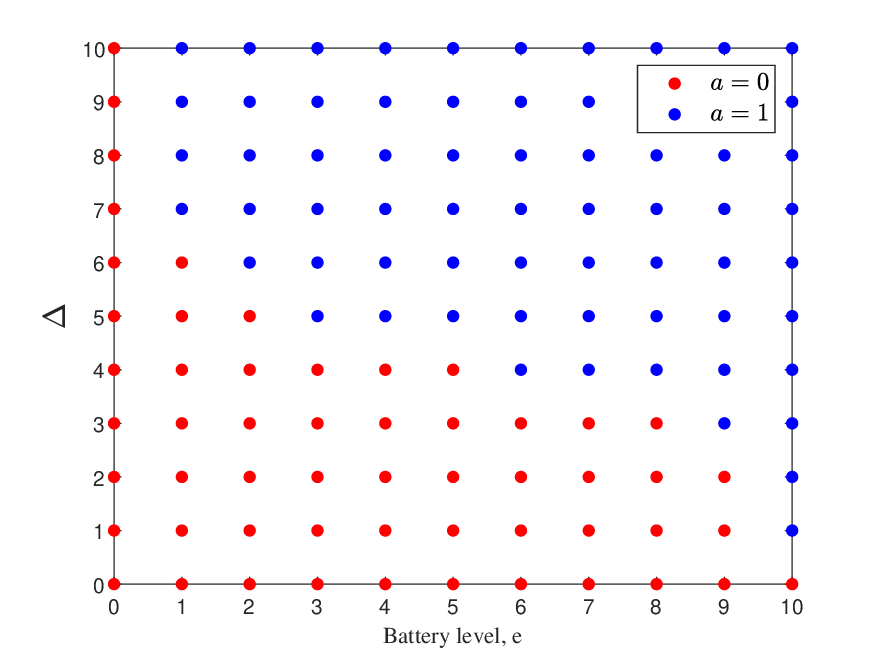}
		}

  \subfigure[$X =0, \beta = 0.4$]{\centering
			\includegraphics[trim=0.5cm 0.05cm 1.1cm 0.6cm,width=0.48\linewidth, clip]{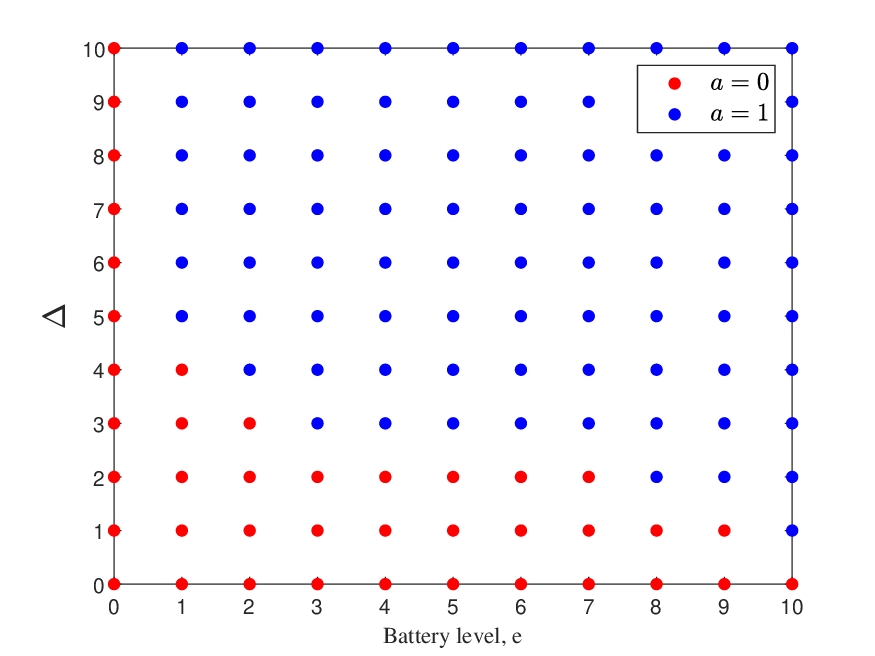}
		}
  \subfigure[$X =1, \beta = 0.4$]{\centering
			\includegraphics[trim=0.5cm 0.05cm 1.1cm 0.6cm,width=0.48\linewidth, clip]{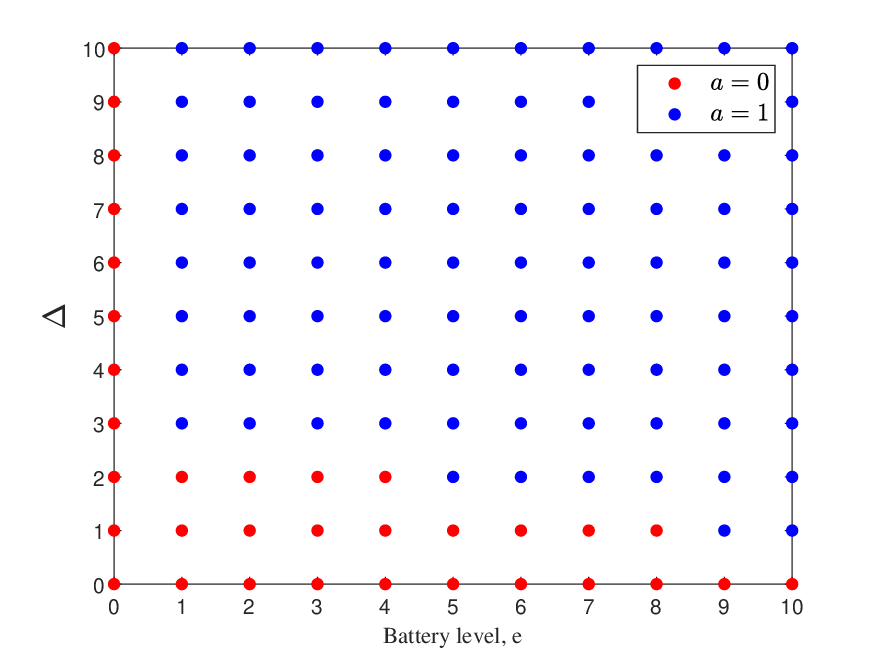}
		}
\caption{The structure of the optimal transmission policy for $p = 0.7$, $q = 0.4$, and $p_{s} = 0.5$.}
\label{VIA_Battery_p0.7q0.4}
	\end{figure}

 Figs. \ref{VIA_Battery_p0.4q0.7} and \ref{VIA_Battery_p0.7q0.4} illustrate the structure of the optimal transmission policy in terms of the VIA, depending on the state of the source for $p_{s} = 0.5$, $E_{\text{max}} = 10$, $\Delta_{\text{max}} = 10$, and selected values of $\beta$, $p$, and $q$. Obviously, when the battery level is empty $(e=0)$ or the VIA is $0$ $(\Delta = 0)$, the optimal transmission policy is $a=0$. Otherwise, the optimal transmission policy follows a threshold-based approach, confirming Theorem \ref{theorem_optimalpolicy}. In this case, if the optimal transmission action for the state $S = (e, X, \Delta)$ is $a=1$, then for the states $S^\prime= (e, X, \Delta^{\prime})$ and $S^{\prime\prime} = (e^{\prime}, X, \Delta)$, where $\Delta^{\prime} > \Delta$ and $e^{\prime}>e$, the action is also $a=1$. Additionally, we observe that when $q>p$, the threshold for taking action $a=1$ is lower for $X = 0$ compared to $X = 1$. Conversely, when $p > q$, the threshold for taking action $a=1$ is higher for $X=0$ than for $X=1$. This occurs because when $q>p$, the source is more likely to be in state $X=0$, whereas when $p>q$, the source is more likely to be in state $X=1$. Therefore, different thresholds for the optimal transmission action apply to different states of the source. In addition, as $\beta$ increases, the energy arrival probability also increases, resulting in the optimal transmission policy being achieved with a lower threshold for the VIA. 
\par
\begin{figure}[!t]
		\centering 
		\subfigure[$\beta = 0.5$]{\includegraphics[trim=0.5cm 0.05cm 1.1cm 0.6cm,width=0.48\linewidth, clip]{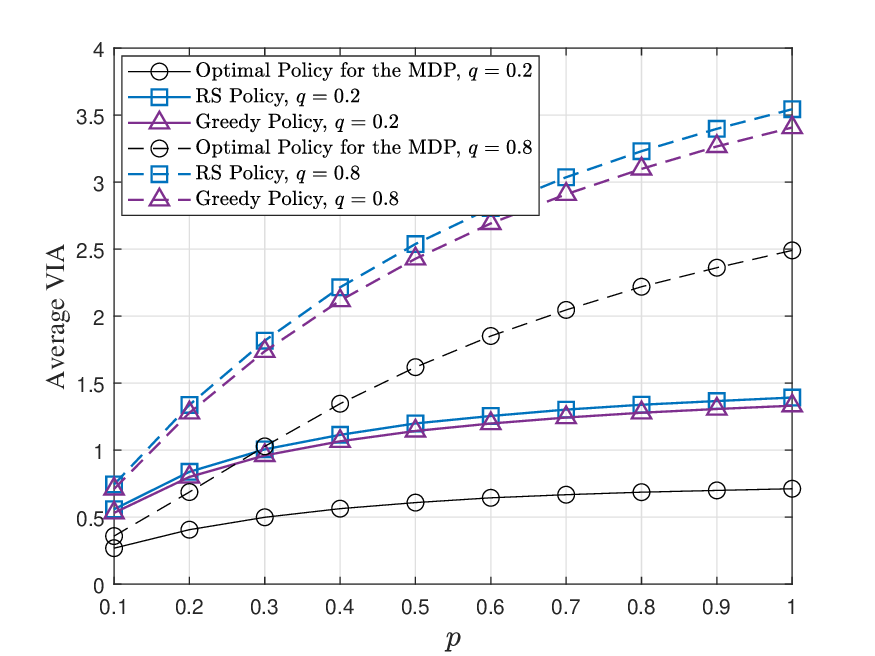}

		}
		\subfigure[$\beta = 0.8$]{\centering
			\includegraphics[trim=0.5cm 0.05cm 1.1cm 0.6cm,width=0.48\linewidth, clip]{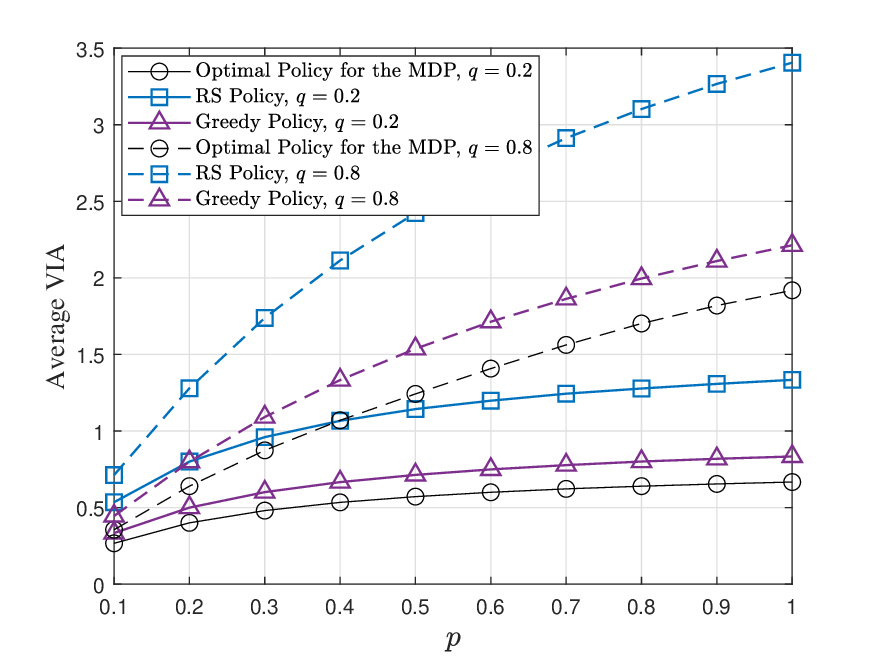}
		}
\caption{Average VIA as a function of $p$ and $q$ for $p_{s} = 0.5$, $E_{\text{max}} = 10$, $\Delta_{\text{max}} = 10$, and selected values of $\beta$.}
\label{AvgVIA_PQ}
\end{figure}

\begin{figure}[!t]
		\centering 
		\subfigure[$\beta = 0.5$]{\includegraphics[trim=0.5cm 0.05cm 1.1cm 0.6cm,width=0.48\linewidth, clip]{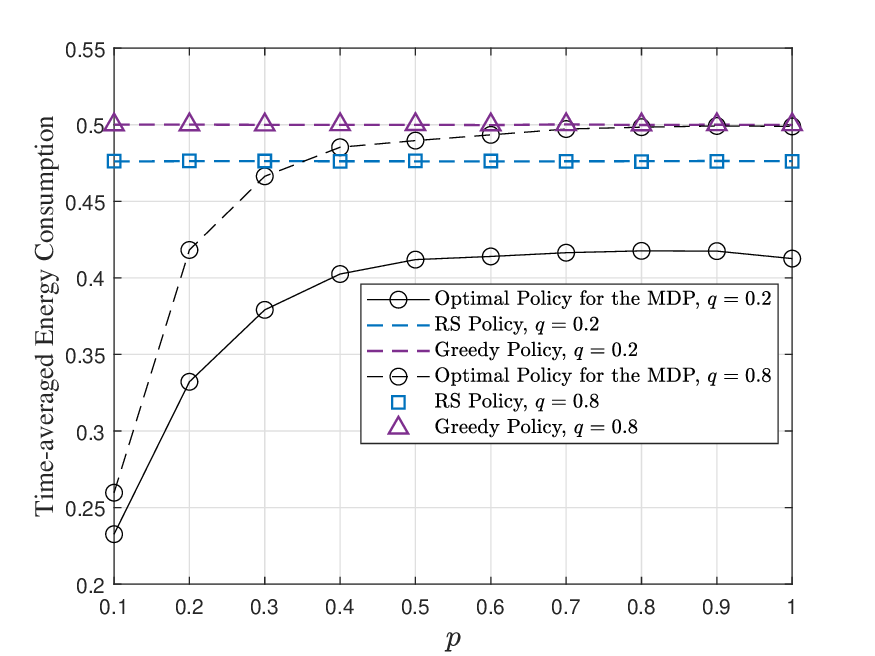}

		}
		\subfigure[$\beta = 0.8$]{\centering
			\includegraphics[trim=0.5cm 0.05cm 1.1cm 0.6cm,width=0.48\linewidth, clip]{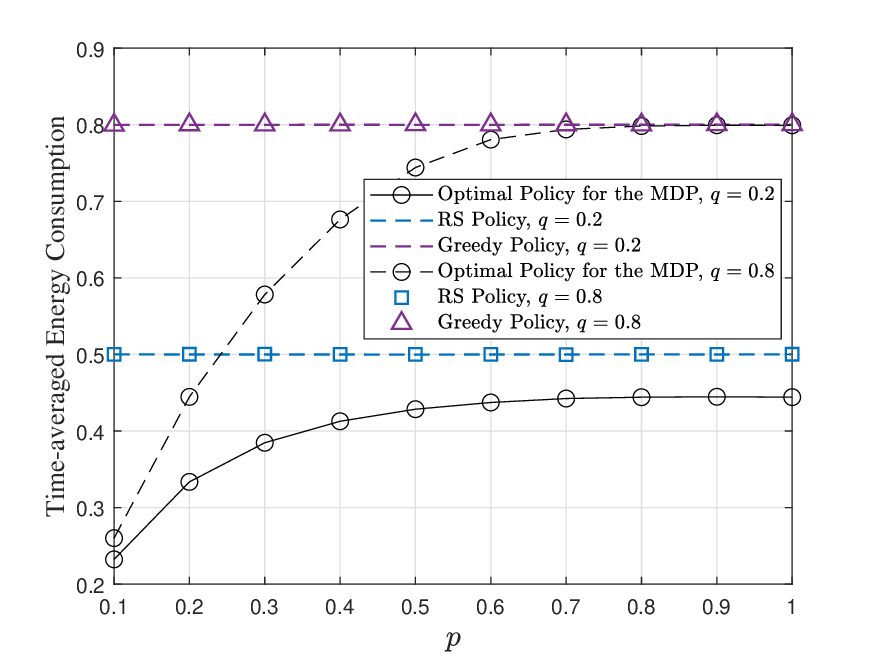}
		}
\caption{Time-average energy consumption as a function of $p$ and $q$ for $p_{s} = 0.5$, $E_{\text{max}} = 10$, $\Delta_{\text{max}} = 10$, and selected values of $\beta$.}
\label{TimeaveragedEnergy}
\end{figure}

\begin{figure}[!t]
		\centering 
		\subfigure[$p_{s} = 0.1$]{\includegraphics[trim=0.5cm 0.05cm 1.1cm 0.6cm,width=0.48\linewidth, clip]{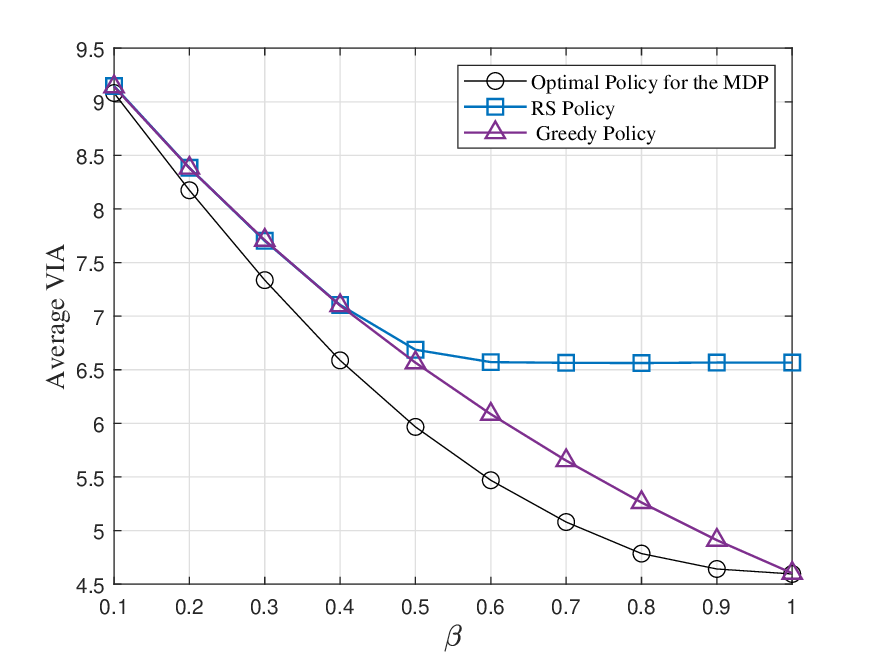}

		}
		\subfigure[$p_{s} = 0.9$]{\centering
			\includegraphics[trim=0.5cm 0.05cm 1.1cm 0.6cm,width=0.48\linewidth, clip]{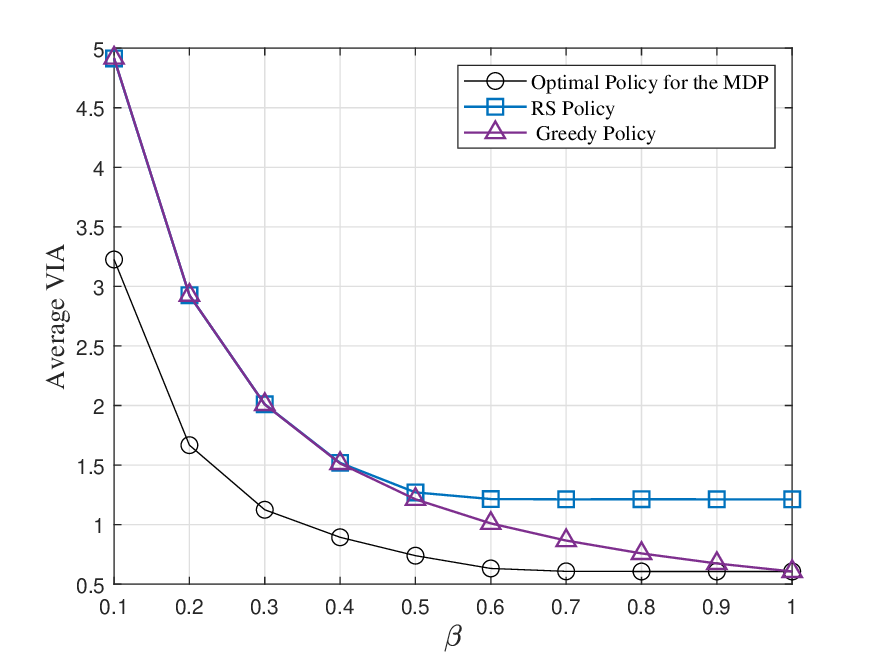}
		}
\caption{Average VIA as a function of $\beta$ for $p = 0.5$, $q = 0.6$, $E_{\text{max}} = 10$, and $\Delta_{\text{max}} = 10$.}
\label{ComparePolicies_beta}
\end{figure}

 The average VIA is depicted in Fig. \ref{AvgVIA_PQ} as a function of $p$, with $p_{s} = 0.5$, $E_{\text{max}} = 10$, and $\Delta_{\text{max}} = 10$, along with the selected values of $q$ and $\beta$. We observe that the average VIA increases when the source evolves rapidly. This is because, under the optimal policy, transmissions are triggered for higher VIA values when the source changes rapidly compared to when it changes slowly. Moreover, the optimal policy outperforms RS and greedy policies for rapid and slow source changes. The reason is that the optimal policy considers the VIA values at each time slot and efficiently utilizes battery energy for transmission to minimize average VIA. In contrast, the other policies transmit without considering the VIA state. For example, in RS and greedy policies, updates may be transmitted even when VIA is $0$, which is unnecessary. In such scenarios, with low battery levels, the transmitter may fail to send updates for higher VIA values due to a depleted battery, resulting in a detrimental effect on the system's performance. Fig. \ref{TimeaveragedEnergy} illustrates the time-averaged energy consumption as a function of $p$ for $p_{s} = 0.5$, $E_{\text{max}} = 10$, $\Delta_{\text{max}} = 10$, and selected values of $q$ and $\beta$. In the RS and greedy policies, we see that time-averaged energy consumption depends on $\beta$ and not on $p$ or $q$. This is because, in the RS policy, the transmission action occurs with probability $p_{\alpha}$ when the battery is not empty; therefore, the time-averaged energy consumption is less than or equal to $\min\{p_{\alpha}, \beta\}$. Furthermore, in the greedy policy, since the transmitter sends updates whenever the battery is not empty, the time-averaged energy consumption is equal to $\beta$. Fig. \ref{ComparePolicies_beta} shows the average VIA as a function of $\beta$ for $p = 0.5$, $q = 0.6$, $E_{\text{max}} = 10$, $\Delta_{\text{max}} = 10$, and selected values of $p_{s}$. This figure demonstrates that the average VIA decreases with increasing $\beta$. This decrease occurs because higher $\beta$ values correspond to increased energy arrivals, reducing constraints on transmission updates.
Additionally, for lower $\beta$ values, the optimal policy outperforms other policies. Effective energy management is crucial for achieving a lower average VIA when $\beta$ is low. Conversely, when $\beta$ is high, the greedy policy converges to the optimal policy. This convergence implies that energy arrives with high probability at each time slot, eliminating energy constraints and necessitating transmission at every slot to optimize performance. 
\section{Conclusion}
\par We studied a time-slotted communication system in which, at each time slot, an EH sensor observes the state of a two-state DTMC and decides whether to transmit status updates to a receiver over a wireless channel. To develop an optimal transmission policy that minimizes the average VIA while accounting for energy limitations, we formulated the problem using the MDP framework to identify the optimal policy. Our results demonstrated that the optimal transmission policy follows a threshold structure based on the battery level, source state, and VIA. Finally, we numerically evaluated the effect of various system parameters on the average VIA and the performance of the optimal policy.
\vspace{-0.3cm}
\appendices
\section{Proof of Theorem {\ref{theorem_optimalpolicy}}}
	\label{Appendix1: theorem1}
 \vspace{-0.15cm}
 \par The Bellman's equation at the state $S = \big(e, X, \Delta\big)$ given in \eqref{BellmanEquation} can be simplified as follows
 \vspace{-.27cm}
 \setlength{\belowdisplayskip}{1pt}
 \begin{align}
      \label{BellmanEquation2}
         \theta^{*}+ V(S) &= \Delta+\underset{a\in \{0,1\}}{\min} \Biggl\{\sum_{S^{\prime}\in \mathfrak{S}}\mathrm{Pr}\big[S^{\prime}|S,a\big]V(S^{\prime})\Biggr\}.
      \end{align}
Now, using \eqref{BellmanEquation2}, the optimal transmission action is obtained as
\vspace{-0.6cm}
\setlength{\belowdisplayskip}{1pt}
\begin{align}
      \label{OptimalPolicy2}
          \!\!\!a^{*}\!(S) \!\in\!  \argmin_{a\in\left\{0,1\right\}}\!\Biggl\{\!\sum_{S^{\prime}\in \mathfrak{S}}\!\!\mathrm{Pr}\!\big[S^{\prime}|S,a\big]\!V\!(S^{\prime})\!\Biggr\}\!\! =\!\! 
          \begin{cases}
              0, &\!\!\!\Delta\! V\!(S)\!\geqslant\! 0,\\
              1,&\!\!\!\Delta\! V\!(S)\!<\!0,
          \end{cases}
      \end{align}
       
      where $\Delta \!V\!(S) \!\!=\!\! V^{1}\!(S)\!\!-\!\!V^{0}\!(S)$, \!\!$V^{1}\!(S)\!\!=\!\!\!  \sum_{S^{\prime}\in \mathfrak{S}}\!\!\mathrm{Pr}\!\big[S^{\prime}|S,a\!=\!1\big]\!V\!(S^{\prime})$, and $V^{0}(S)= \sum_{S^{\prime}\in \mathfrak{S}}\mathrm{Pr}\big[S^{\prime}|S,a=0\big]V(S^{\prime})$. When $e\!=\!0$ and $\Delta\! \geqslant\! 0$, it can be easily shown that $\Delta\! V\!(S) \!=\! 0$. Therefore, when the battery level is empty, the optimal transmission action is $a = 0$. Now, we investigate the other case where $e \neq 0$ and $\Delta \geqslant 0$. For that, we consider the scenario where the state of the source is $0$, i.e., $X=0$; a similar proof can be provided for $X=1$. We first consider the case where $\Delta = 0$. According to \eqref{VIA}, when $\Delta=0$, the sole factor influencing the increase in $\Delta$ is the source changes. In this scenario, the transmission of status updates, whether successful or not, does not affect $\Delta$. Therefore, since transmission solely consume energy without impacting $\Delta$, the optimal transmission action in this case is $a = 0$. Now, we consider the case where $\Delta>0$ and define the state $S = (e, 0, \Delta)$.  For this state, using \eqref{TransProb_001i} and \eqref{TranProb_ED}, we can obtain $V^{0}(S), V^{1}(S)$, and $\Delta V(S)$ as follows:
      \vspace{-0.15cm}
\begin{align}
    \!\!V^{0}(e, 0, \Delta) &\!=\!\Bb (1\!-\!p)V(e,0,\Delta)\!+\!\beta p V(e+1,1,\Delta+1)\notag\\
    &\!+\!\beta(1\!-\!p)V(e+1,0,\Delta)
    \!+\!\Bb pV(e,1,\Delta\!+\!1),\label{V0SD}\\
    \hspace{-1.6cm}V^{1}(e, 0, \Delta)&=\beta(1\!-\!p)p_{s}V(e,0,0)+\Bb(1\!-\!p)p_{s}V(e\!-\!1,0,0)\notag\\
    &+\Bb(1\!-\!p)p_{f}V(e\!-\!1,0,\Delta)
    +\beta pp_{f}V(e,1,\Delta\!+\!1)\notag\\
    &+\Bb p p_{s}V(e\!-\!1,1,1)+\beta(1\!-\!p)p_{f} V(e,0,\Delta)\notag\\
    &+\beta p p_{s}V(e,1,1)+\Bb pp_{f}V(e\!-\!1,1,\Delta\!+\!1),\label{V1SD}\\
    &\hspace{-1.6cm}\Delta V(e,0,\Delta) = V^{1}(S)-V^{0}(S)= \notag\\
    & \hspace{-1.6cm}\beta(1\!-\!p)\Big[p_{s}V(e,0,0)\!+\!p_{f}V(e,0,\Delta)\!-\!V(e\!+\!1,0,\Delta)\Big]\notag\\
    &\hspace{-1.7cm}+\!\Bb(1\!-\!p)\Big[p_{s}V(e\!-\!1,0,0)\!+\!p_{f}V(e\!-\!1,0,\Delta)\!-\!V(e,0,\Delta)\Big]\notag\\
   &\hspace{-1.7cm}+\!\beta p \Big[p_{s}V(e,1,1)\!+\!p_{f}V(e,1,\Delta\!+\!1)-V(e\!+\!1,1,\Delta\!+\!1)\Big]\notag\\
  &\hspace{-1.7cm}+\!\!\Bb p\!\Big[p_{s}V\!(e\!\!-\!\!1,1,1)\!\!+\!\!p_{f}\!V\!(e\!\!-\!\!1,1,\Delta\!+\!1)\!\!-\!\!V\!(e,1,\Delta\!+\!1)\Big]. \label{DV_D}
\end{align}
In what follows, we demonstrate that $\Delta V(e, 0, \Delta)$ is a decreasing function of $\Delta$. Therefore, $\Delta V(e, 0, \Delta)$ can become negative for a large value of $\Delta$, resulting in the action $a = 1$ for $\Delta\geqslant \Delta_{\text{TH}}$. To prove that $\Delta V(e, 0, \Delta)$ decreases with respect to $\Delta$, we define two states $S = (e, 0, \Delta^{+})$ and $S = (e, 0, \Delta^{-})$ where $\Delta^{+} \geqslant \Delta^{-}$, and show that $\Delta V(e, 0, \Delta^{+}) - \Delta V(e, 0, \Delta^{-}) \leqslant 0$. Here, we assume that $S_{e+x,0} = (e+x, 0, \Delta)$,$ 
\Tilde{S}_{e+x,1} = (e+x, 1, \Delta+1)$, $S^{+}_{e+x,0} = (e+x, 0, \Delta^{+}), S^{-}_{e+x,0} = (e+x, 0, \Delta^{-}), 
\Tilde{S}^{+}_{e+x,1} = (e+x, 1, \Delta^{+}+1), \Tilde{S}^{-}_{e+x,1} \!=\! (e\!+\!x, 1, \Delta^{-}\!+\!1)
$, where $x\!\in\! \mathbb{ Z}$. Now, using \eqref{DV_D}, we can write 
\vspace{-0.3cm}
\begin{align}
\label{DV_plus_minus}
    &\Delta V(S^{+}_{e,0})-\Delta V(S^{-}_{e,0}) \notag\\
    &= \Bb (1-p)\bigg\{p_{f}\Big[V(S^{+}_{e-1,0})-V(S^{-}_{e-1,0})\Big]\notag\\&\!-\!\Big[V(S^{+}_{e,0})\!-\!V(S^{-}_{e,0})\Big]\!\bigg\}\!\!+\!\!\beta (1\!-\!p)\bigg\{\!p_{f}\Big[V(S^{+}_{e,0})\!-\!V(S^{-}_{e,0})\Big]\notag\\
    &\!-\!\Big[V(S^{+}_{e+1,0})\!-\!V(S^{-}_{e+1,0})\Big]\!\bigg\}\!\!+\!\!\beta p\bigg\{\!p_{f}\Big[V(\Tilde{S}^{+}_{e,1})\!-\!V(\Tilde{S}^{-}_{e,1})\Big]\notag\\
    &\!-\!\Big[V\!(\Tilde{S}^{+}_{e+1,1})\!-\!V\!(\Tilde{S}^{-}_{e+1,1})\Big]\!\bigg\}\!\!+\!\!\Bb p\bigg\{\!p_{f}\Big[V\!(\Tilde{S}^{+}_{e-1,1})\!-\!V\!(\Tilde{S}^{-}_{e-1,1})\Big]\notag\\
    &-\Big[V(\Tilde{S}^{+}_{e,1})-V(\Tilde{S}^{-}_{e,1})\Big]\bigg\}.
\end{align}
\vspace{-0.1cm}
According to \eqref{DV_plus_minus}, to demonstrate that $\Delta V(S^{+}_{e,0}) - \Delta V(S^{-}_{e,0}) \leqslant 0$, it suffices to prove that $p_{f}\Big[V(S^{+}_{e-1,0}) - V(S^{-}_{e-1,0})\Big] - \Big[V(S^{+}_{e,0}) - V(S^{-}_{e,0})\Big] \leqslant 0$. We proceed with the value iteration algorithm and mathematical induction for the proof. The value iteration algorithm converges to the value function of Bellman’s equation regardless of the initial value assigned to $V_{0}(S)$, i.e, $\lim_{k\to \infty} V_{k}(S), \forall S \in \mathfrak{S}$. Therefore, it is sufficient to establish the following inequality for all $t \in \{0, 1, 2, \ldots\}$:
\vspace{-0.2cm}
\begin{align}
\label{condition_DV_dec}
    \!\!p_{f}\Big[V_{t}(S^{+}_{e-1,0})\!-\!V_{t}(S^{-}_{e-1,0})\Big]\!\!-\!\!\Big[V_{t}(S^{+}_{e,0})\!-\!V_{t}(S^{-}_{e,0})\Big]\!\leqslant\! 0.
\end{align}
We first suppose that $V_{0}(S) = 0, \forall S \in \mathfrak{S}$, therefore, \eqref{condition_DV_dec} holds for $t=0$. Now, we extend this assumption to $t>0$ and verify whether it holds for $t+1$. By defining $V^{0}_{t+1}(S)=  \sum_{S^{\prime}\in \mathfrak{S}}\mathrm{Pr}\big[S^{\prime}|S,a=0\big]V_{t}(S^{\prime})$, and $V^{1}_{t+1}(S)= \sum_{S^{\prime}\in \mathfrak{S}}\mathrm{Pr}\big[S^{\prime}|S,a=1\big]V_{t}(S^{\prime})$, and using \eqref{TransProb_001i} and \eqref{TranProb_ED}, we can obtain $V^{0}_{t+1}(S)$ and $V^{1}_{t+1}(S)$ for $S = (e, 0, \Delta)$, where $e\geqslant 0$ and $\Delta>0$ as follows
\vspace{-.2cm}
\begin{align}
\label{VS01_T}
     V^{0}_{t+1}(S_{e,0}) &= \Bb (1-p)V_{t}(S_{e,0})+\Bb pV_{t}(\Tilde{S}_{e,1})\notag\\
     &+\beta(1-p)V_{t}(S_{e+1,0})+\beta p V_{t}(\Tilde{S}_{e+1,1}),\notag\\
    \hspace{-1.6cm}V^{1}_{t+1}(S_{e,0})&=\beta(1\!-\!p)p_{s}V_{t}(e,0,0)\!+\!\Bb(1\!-\!p)p_{s}V_{t}(e\!-\!1,0,0)\notag\\
    &+\beta(1-p)p_{f}V_{t}(S_{e,0})\!+\!\Bb(1\!-\!p)p_{f}V_{t}(S_{e-1,0})\notag\\
    &+\beta p p_{s}V_{t}(e,1,1)+\Bb p p_{s}V_{t}(e-1,1,1)\notag\\
    &+\beta pp_{f}V_{t}(\Tilde{S}_{e,1})+\Bb p p_{f}V_{t}(\Tilde{S}_{e-1,1}).
\end{align}
Now, using \eqref{VS01_T}, the value iteration algorithm is given by
\vspace{-0.1cm}
\begin{align}
\label{VIA_SE0D}
V_{t+1}(S_{e,0}) = \Delta+\min\Big\{V^{0}_{t+1}(S_{e,0}),V^{1}_{t+1}(S_{e,0})\Big\}.    
\end{align}
Using \eqref{VIA_SE0D}, the condition in \eqref{condition_DV_dec} for time slot $t+1$ can be written as
\vspace{-0.2cm}
\begin{align}
\label{condition_DV_dec_T}
&\!\!\!\!p_{f}\!\Big[V_{t+1}(S^{+}_{e-1,0})\!\!-\!\!V_{t+1}(S^{-}_{e-1,0})\Big]\!\!-\!\Big[\!V_{t+1}(S^{+}_{e,0})\!\!-\!\!V_{t+1}(S^{-}_{e,0})\Big]\notag\\
&\!\!\!\!=\!\!-p_{s}(\Delta^{+}\!\!-\!\!\Delta^{-})\!+\!p_{f}\bigg[\!\min\Big\{V^{0}_{t+1}(S^{+}_{e-1,0}),V^{1}_{t+1}(S^{+}_{e-1,0})\Big\}\notag\\
&\!\!\!-\min\Big\{V^{0}_{t+1}(S^{-}_{e-1,0}),V^{1}_{t+1}(S^{-}_{e-1,0})\Big\}\!\bigg]
    \notag\\
    &\!\!\!-\!\!\bigg[\!\!\min\!\Big\{\!V^{0}_{t+1}(S^{+}_{e,0}),V^{1}_{t+1}(S^{+}_{e,0})\!\Big\}\notag\\
    &\!\!-\min\!\Big\{\!V^{0}_{t+1}(S^{-}_{e,0}),V^{1}_{t+1}(S^{-}_{e,0})\!\Big\}  \!\bigg]\!\!\leqslant\!\! 0.
\end{align}
The first term of \eqref{condition_DV_dec_T} is negative. To prove that the other terms are also negative, we consider four cases. In Case 1, we assume that $V^{0}_{t+1}(S^{-}_{e-1,0})\leqslant V^{1}_{t+1}(S^{-}_{e-1,0})$ and $V^{0}_{t+1}(S^{+}_{e,0})\leqslant V^{1}_{t+1}(S^{+}_{e,0})$. In Case 2, we assume $V^{0}_{t+1}(S^{-}_{e-1,0})\leqslant V^{1}_{t+1}(S^{-}_{e-1,0})$ and $V^{0}_{t+1}(S^{+}_{e,0})>V^{1}_{t+1}(S^{+}_{e,0})$. In Case 3, we consider the scenario where$V^{0}_{t+1}(S^{-}_{e-1,0})> V^{1}_{t+1}(S^{-}_{e-1,0})$ and $V^{0}_{t+1}(S^{+}_{e,0})\leqslant V^{1}_{t+1}(S^{+}_{e,0})$. In Case 4, we assume that $V^{0}_{t+1}(S^{-}_{e-1,0})> V^{1}_{t+1}(S^{-}_{e-1,0})$ and $V^{0}_{t+1}(S^{+}_{e,0})> V^{1}_{t+1}(S^{+}_{e,0})$. In what follows, we provide the proof for Case 1, and a similar approach can be applied to prove the remaining cases. For the Case 1, the expression given in \eqref{condition_DV_dec_T} simplifies to
\vspace{-0.3cm}
    \begin{align}
    \label{condition_case1_expr1}
        &\!\!\!\!\!\!p_{f}\bigg[\!\min\Big\{\!V^{0}_{t+1}(S^{+}_{e-1,0}),V^{1}_{t+1}(S^{+}_{e-1,0})\!\Big\}\!-\!V^{0}_{t+1}(S^{-}_{e-1,0})\bigg]\notag\\
    &\!\!\!\!\!-\bigg[V^{0}_{t+1}(S^{+}_{e,0})-\min\Big\{V^{0}_{t+1}(S^{-}_{e,0}),V^{1}_{t+1}(S^{-}_{e,0})\Big\}  \bigg]\leqslant 0.
    \end{align}
  Using $\min\{x,y\} = x + \min\{0,y-x\}$, we can simplify \eqref{condition_case1_expr1} as follows:
  \vspace{-0.1cm}
\begin{align}
\label{condition_case1_expr2}
    &p_{f}\!\bigg[\!V^{0}_{t+1}(S^{+}_{e-1,0})\!-\!V^{0}_{t+1}(S^{-}_{e-1,0})\!\bigg]
    \!\!-\!\!\bigg[\!V^{0}_{t+1}(S^{+}_{e,0})-V^{0}_{t+1}(S^{-}_{e,0})\!\bigg]\notag\\
    &+p_{f}\underbrace{\min\Big\{0, V^{1}_{t+1}(S^{+}_{e-1,0})-V^{0}_{t+1}(S^{+}_{e-1,0})\Big\}}_{\leqslant 0}\notag\\
    &
    +\underbrace{\min\Big\{0, V^{1}_{t+1}(S^{-}_{e,0})-V^{0}_{t+1}(S^{-}_{e,0})\Big\}}_{\leqslant 0}\leqslant 0.
\end{align}
Since the second term in \eqref{condition_case1_expr2} is negative, it suffices to demonstrate that
\setlength{\belowdisplayskip}{-1.8pt}
\vspace{-0.2cm}
\begin{align}
\label{condition_case1_expr3}
    \!\!\!p_{f}\bigg[\!V^{0}_{t+1}\!(S^{+}_{e-1,0})\!\!-\!\!V^{0}_{t+1}\!(S^{-}_{e-1,0})\!\bigg]\!\!-\!\!\bigg[\!V^{0}_{t+1}\!(S^{+}_{e,0})\!-\!V^{0}_{t+1}\!(S^{-}_{e,0})\bigg]\!\!\leqslant\! 0.
\end{align}
\vspace{-0.3cm}
 
Now, using \eqref{VS01_T}, we can write \eqref{condition_case1_expr3} as follows
\vspace{-0.2cm}
{\setlength{\jot}{1pt}
\begin{align}
\label{condition_case1_expr4}
 &\Bb (1-p)\bigg\{p_{f}\Big[V(S^{+}_{e-1,0})-V(S^{-}_{e-1,0})\Big]\notag\\
 \vspace{-0.3cm}
 &\!-\!\Big[\!V(S^{+}_{e,0})\!-\!V(S^{-}_{e,0})\!\Big]\bigg\}\!+\!\beta (1\!-\!p)\bigg\{\!p_{f}\Big[\!V(S^{+}_{e,0})\!-\!V(S^{-}_{e,0})\!\Big]\notag\\
 &\!-\!\Big[\!V(S^{+}_{e+1,0})\!-\!V(S^{-}_{e+1,0})\!\Big]\!\bigg\}\!+\!\beta p\bigg\{\!\Bb\Big[\!V(\Tilde{S}^{+}_{e,1})\!-\!V(\Tilde{S}^{-}_{e,1})\!\Big]\notag\\
 &-\!\!\Big[\!V\!(\Tilde{S}^{+}_{e+1,1})\!\!-\!\!V\!(\Tilde{S}^{-}_{e+1,1})\!\Big]\!\bigg)\!\bigg\}\!+\!\Bb p\bigg\{\!p_{f}\Big[V\!(\Tilde{S}^{+}_{e-1,1})\!-\!V\!(\Tilde{S}^{-}_{e-1,1})\Big]\notag\\
    &-\Big[V(\Tilde{S}^{+}_{e,1})-V(\Tilde{S}^{-}_{e,1})\Big]\bigg\}\leqslant 0. 
\end{align}}
Using the assumption given in \eqref{condition_DV_dec}, we conclude that the expression in \eqref{condition_case1_expr4} is negative. This confirms that the proof is valid for the specific condition described in Case 1.
\vspace{-0.2cm}
\bibliographystyle{IEEEtran}
\bibliography{ref}

\begin{thebibliography}{10}
\providecommand{\url}[1]{#1}
\csname url@samestyle\endcsname
\providecommand{\newblock}{\relax}
\providecommand{\bibinfo}[2]{#2}
\providecommand{\BIBentrySTDinterwordspacing}{\spaceskip=0pt\relax}
\providecommand{\BIBentryALTinterwordstretchfactor}{4}
\providecommand{\BIBentryALTinterwordspacing}{\spaceskip=\fontdimen2\font plus
\BIBentryALTinterwordstretchfactor\fontdimen3\font minus \fontdimen4\font\relax}
\providecommand{\BIBforeignlanguage}[2]{{%
\expandafter\ifx\csname l@#1\endcsname\relax
\typeout{** WARNING: IEEEtran.bst: No hyphenation pattern has been}%
\typeout{** loaded for the language `#1'. Using the pattern for}%
\typeout{** the default language instead.}%
\else
\language=\csname l@#1\endcsname
\fi
#2}}
\providecommand{\BIBdecl}{\relax}
\BIBdecl

\bibitem{abd2019role}
M.~A. Abd-Elmagid, N.~Pappas, and H.~S. Dhillon, ``{On the role of age of information in the {Internet of Things}},'' \emph{IEEE Communications Magazine}, vol.~57, no.~12, pp. 72--77, 2019.

\bibitem{shreedhar2019age}
T.~Shreedhar, S.~K. Kaul, and R.~D. Yates, ``{An age control transport protocol for delivering fresh updates in the {Internet-of-Things}},'' in \emph{IEEE 20th International Symposium on WoWMoM}, 2019, pp. 1--7.

\bibitem{kountouris2021semantics}
M.~Kountouris and N.~Pappas, ``{Semantics-empowered communication for networked intelligent systems},'' \emph{IEEE Commun. Mag.}, 2021.

\bibitem{popovski2020semantic}
P.~Popovski, O.~Simeone, F.~Boccardi, D.~G{\"u}nd{\"u}z, and O.~Sahin, ``{Semantic-effectiveness filtering and control for post-5G wireless connectivity},'' \emph{Journal of the Indian Institute of Science}, 2020.

\bibitem{popovski2022perspective}
P.~Popovski, F.~Chiariotti, K.~Huang, A.~E. Kal{\o}r, M.~Kountouris, N.~Pappas, and B.~Soret, ``{A perspective on time toward wireless 6G},'' \emph{Proceedings of the IEEE}, vol. 110, no.~8, pp. 1116--1146, 2022.

\bibitem{kaul2012real}
S.~Kaul, R.~Yates, and M.~Gruteser, ``Real-time status: How often should one update?'' in \emph{2012 Proceedings IEEE INFOCOM}, 2012.

\bibitem{maatouk2020age}
A.~Maatouk, S.~Kriouile, M.~Assaad, and A.~Ephremides, ``{The age of incorrect information: A new performance metric for status updates},'' \emph{IEEE/ACM Transactions on Networking}, 2020.

\bibitem{pappas2021goal}
N.~Pappas and M.~Kountouris, ``{Goal-oriented communication for real-time tracking in autonomous systems},'' in \emph{IEEE ICAS}, 2021.

\bibitem{MSalimnejadTCOM2024}
M.~Salimnejad, M.~Kountouris, and N.~Pappas, ``{Real-time Reconstruction of Markov Sources and Remote Actuation over Wireless Channels},'' \emph{IEEE Transactions on Communications}, 2024.

\bibitem{yates2021age}
R.~D. Yates, ``{The age of gossip in networks},'' in \emph{IEEE ISIT}, 2021.

\bibitem{salimnejad2024age}
M.~Salimnejad, M.~Kountouris, A.~Ephremides, and N.~Pappas, ``{Age of Information Versions: a Semantic View of Markov Source Monitoring},'' \emph{arXiv preprint arXiv:2406.14594}, 2024.

\bibitem{Yates2015}
R.~D. Yates, ``{Lazy is timely: Status updates by an energy harvesting source},'' in \emph{IEEE ISIT}, 2015.

\bibitem{wu2017optimal}
X.~Wu, J.~Yang, and J.~Wu, ``{Optimal status update for age of information minimization with an energy harvesting source},'' \emph{IEEE Transactions on Green Communications and Networking}, 2017.

\bibitem{arafa2019age}
A.~Arafa, J.~Yang, S.~Ulukus, and H.~V. Poor, ``{Age-minimal transmission for energy harvesting sensors with finite batteries: Online policies},'' \emph{IEEE Transactions on Information Theory}, vol.~66, no.~1, pp. 534--556, 2019.

\bibitem{abd2019online}
M.~A. Abd-Elmagid, H.~S. Dhillon, and N.~Pappas, ``{Online age-minimal sampling policy for RF-powered IoT networks},'' in \emph{2019 IEEE Global Communications Conference (GLOBECOM)}, 2019, pp. 1--6.

\bibitem{Stamatakis2019}
G.~Stamatakis, N.~Pappas, and A.~Traganitis, ``{Control of Status Updates for Energy Harvesting Devices That Monitor Processes with Alarms},'' in \emph{IEEE Globecom Workshops (GC Wkshps)}, 2019.

\bibitem{hatami2022demand}
M.~Hatami, M.~Leinonen, Z.~Chen, N.~Pappas, and M.~Codreanu, ``On-demand aoi minimization in resource-constrained cache-enabled iot networks with energy harvesting sensors,'' \emph{IEEE Transactions on Communications}, vol.~70, no.~11, pp. 7446--7463, 2022.

\bibitem{DelfaniWiopt2023}
E.~Delfani and N.~Pappas, ``{Version Age-Optimal Cached Status Updates in a Gossiping Network with Energy Harvesting Sensor},'' in \emph{WiOpt}, 2023.

\bibitem{bertsekas2011dynamic}
D.~P. Bertsekas, \emph{{Dynamic Programming and Optimal Control, Vol. II}}, 3rd~ed.\hskip 1em plus 0.5em minus 0.4em\relax Athena Scientific, 2007.

\end{thebibliography}
\end{document}